%% file: arxiv.tex
\documentclass[aps,prl,reprint,twocolumn,showpacs,floatfix,superscriptaddress]{revtex4-2}

\usepackage{multirow,times}
\usepackage{colortbl}
\usepackage{amssymb,amsmath,amstext}
\usepackage{graphicx}
\usepackage{epstopdf}
\usepackage{color}
\usepackage{bm}
\usepackage{appendix}
\usepackage[utf8]{inputenc}
\usepackage[T1]{fontenc}
\usepackage{dsfont}
\usepackage[normalem]{ulem}
\usepackage{booktabs}
\usepackage[shortlabels]{enumitem}
\usepackage{verbatim}
\usepackage{latexsym}
\usepackage{xcolor}
\usepackage[table]{xcolor}
\usepackage{tabularray}
\usepackage{subcaption}
\usepackage{braket}
\usepackage{ulem}
\definecolor{lblue}{RGB}{51,71,158}
\usepackage[colorlinks=true,citecolor=blue,linkcolor=blue,urlcolor=lblue]{hyperref}
\DeclareMathOperator{\Tr}{Tr}

\makeatletter\AtBeginDocument{\let\@elt\relax}\makeatother

\begin{document}

\title{Unconventional Thermalization of a Localized Chain Interacting with an Ergodic Bath}

\author{Konrad Pawlik}
\email{konrad.pawlik@doctoral.uj.edu.pl}
\affiliation{Szkoła Doktorska Nauk Ścisłych i Przyrodniczych, Uniwersytet Jagielloński, ulica Stanisława \L{}ojasiewicza 11, PL-30-348 Krak\'ow, Poland}
\affiliation{Instytut Fizyki Teoretycznej,  Wydział Fizyki, Astronomii i Informatyki Stosowanej, Uniwersytet Jagielloński, ulica Stanisława \L{}ojasiewicza 11, PL-30-348 Krak\'ow, Poland }

\author{Nicolas Laflorencie}
\email{nicolas.laflorencie@cnrs.fr}
\affiliation{Univ. Toulouse, CNRS, Laboratoire de Physique Th\'eorique, Toulouse, France}
\author{Jakub Zakrzewski}
\email{jakub.zakrzewski@uj.edu.pl}
\affiliation{Instytut Fizyki Teoretycznej,  Wydział Fizyki, Astronomii i Informatyki Stosowanej, Uniwersytet Jagielloński, ulica Stanisława \L{}ojasiewicza 11, PL-30-348 Krak\'ow, Poland }
\affiliation{Mark Kac Complex
    Systems Research Center, Jagiellonian University in Krakow, PL-30-348 Krak\'ow,
    Poland. }

\date{\today}

\begin{abstract}
    The study of many-body localized (MBL) phases intrinsically links spectral properties with eigenstate characteristics: localized systems exhibit Poisson level statistics and area-law entanglement entropy, while ergodic systems display volume-law entanglement and follow random matrix theory predictions, including level repulsion. Here, we introduce the interacting Anderson Quantum Sun model, which significantly deviates from these conventional expectations. In addition to standard localized and ergodic phases, we identify a regime that exhibits volume-law entanglement coexisting with intermediate spectral statistics.
    We also identify another nonstandard regime marked by Poisson level statistics, sub-volume entanglement growth, and rare-event-dominated correlations, indicative of emerging ergodic instabilities. These results highlight unconventional routes of ergodicity breaking and offer fresh perspectives on how Anderson localization may be destabilized.
\end{abstract}

\maketitle

\paragraph{Introduction.}
Many-body localization (MBL)~\cite{Gornyi05, Basko06, Oganesyan07, Pal10}, a spectacular counterexample to the once-dominating belief of the prevalence of ergodic dynamics in many-body interacting systems, has been extensively studied for twenty years already  (see recent reviews for details \cite{Nandkishore15,Alet18,Abanin19,Sierant25}). Typically in finite one-dimensional systems, one may observe both the ergodic phase described by the eigenstate thermalization hypothesis (ETH)~\cite{Deutsch91, Srednicki94, Rigol08, Dalessio16} for small disorder and the MBL phase with the phenomenological description provided by quasi-local integrals of motion (lioms)~\cite{Serbyn13b, Huse14,Ros15, Thomson18} for large disorder. The MBL system fails to thermalize and retains the memory of its initial state~\cite{Pal10, Huse14, Luitz15}, which is accompanied by a suppression of transport~\cite{Nandkishore15, Znidaric16} and a slow, logarithmic in time spread of the entanglement~\cite{DeChiara06, Znidaric08, Bardarson12,Serbyn13a}. The signatures of MBL were observed experimentally in relatively small systems \cite{Schreiber15,Bordia16,Smith16,Luschen17,Leonard23} and at finite times, while mathematical arguments for its existence were also provided \cite{Imbrie16} (for recent advances, see also Refs.~\cite{yin2024,deroeck2024,baldwin25}).

This comfortable picture has been shaken~\cite{Suntajs20e, Weiner19} by others who stressed a strong dependence of the ergodic-MBL crossover on the system size. In effect, the existence
of the MBL phase in the thermodynamic limit remains an open question~\cite{Sierant20b, Panda20, Sels20, Kiefer20, Sierant20p, Abanin21, Kiefer21u, Ghosh22, Sels21,Sierant22,Sels21a, Morningstar22,Laflorencie20, Suntajs20,Luitz20,Sierant23f,Szoldra23} as also summarized in a recent review \cite{Sierant25}.

Another approach, originating from the
so-called avalanche scenario~\cite{DeRoeck17}, has aimed to understand how small thermal regions (appearing, e.g., due to rare Griffiths regions of relatively weak disorder) may spread, possibly destabilizing the MBL phase. This inspired a series of effective models that emulate the influence of such thermal baths on localization~\cite{Luitz17,Potirniche18,Crowley20,Crowley22,Brighi22l,Brighi22,Brighi23,Sierant23,Falcao23,Szoldra24,Colmenarez24,Berger24,Sajid25}. Among those, the quantum sun (QS) model \cite{Luitz17,Suntajs22}
distinguished itself as a toy model with remarkable properties.  In particular, it was shown to exhibit an ergodicity-breaking transition with negligible finite-size effects~\cite{Suntajs22} compared to paradigmatic many-body interacting models such as, e.g., the disordered Heisenberg chain, resulting in better controlled numerics. Subsequent works revealed further notable features of the QS, including numerical evidence of the many-body mobility edge~\cite{Pawlik24} despite earlier arguments against its existence~\cite{DeRoeck16}, identification of Fock space localized phase, multifractal critical point across the parameter space~\cite{Suntajs24s} and the demonstration of the fading ergodicity ansatz~\cite{Kliczkowski24,Swietek25}.

\begin{figure}[t!]
    \centering
    \includegraphics[width=\linewidth]{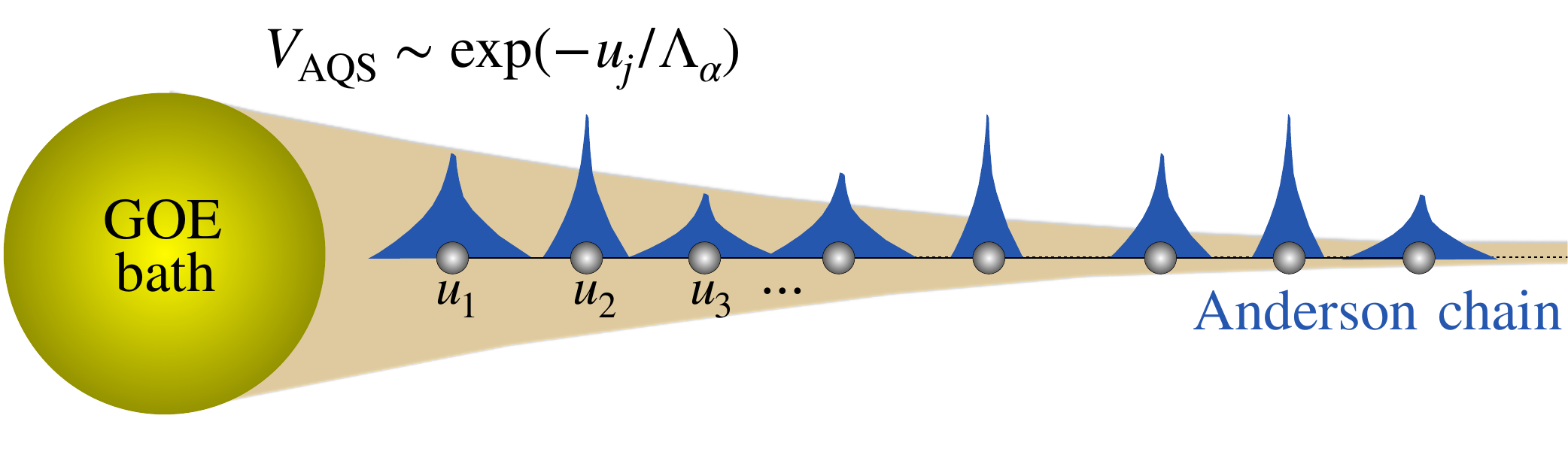}
    \caption{Schematic of the AQS model Eq.~\eqref{eq:model_hamiltonian}. An ergodic bath is coupled to an Anderson chain via exponentially decaying interactions.}
    \label{fig:sketch}
\end{figure}

\begin{figure}[t!]
    \centering
    \includegraphics[width=\linewidth]{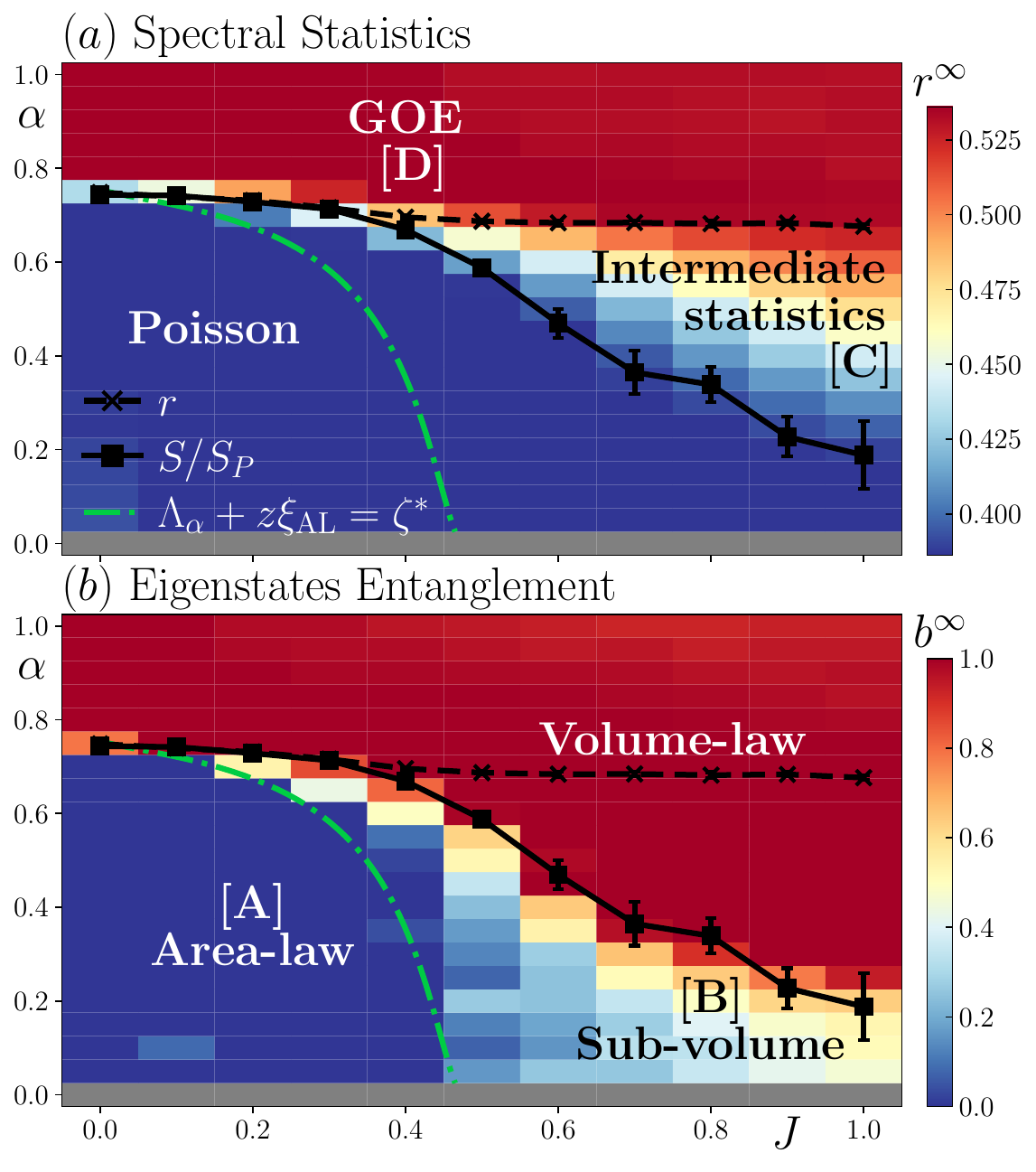}
    \caption{Phase diagram of the AQS model. (a): gap ratio extrapolated to the thermodynamic limit $r^\infty$. (b): power in the power-law dependence of entropy growth $S\propto L^b$ extrapolated to the thermodynamic limit $b^\infty$. Curves on both plots represent spectral (dashed black, $\alpha^{\rm c}_{r,\infty}$) and eigenstate (solid black, $\alpha^{\rm c}_{S,\infty}$) phase transition, regimes of the model are labeled with [A-D]. The dotted-dashed green analytical line indicates the onset of ergodic instabilities. The Anderson limit $\alpha=0$ is excluded and shown in gray.}
    \label{fig:phase_diagram}
\end{figure}

\paragraph{Main results.}
Since the QS has proven to be a good toy model to mimic the MBL transition under the influence of an ergodic bath~\cite{DeRoeck17,Luitz17,Suntajs22,Pawlik24,Kliczkowski24,Swietek25}, it seems well-suited to answer one of the most fundamental questions in the field~\cite{Fleishman80,Gornyi05,Basko06}: How can interactions with an ergodic bath destabilize an Anderson insulator, and via what mechanism(s)\,? To address this question, we consider the QS model, i.e., a spin chain with the interaction of individual spins to the ergodic bath decaying exponentially with the distance, enriched by U(1) nearest neighbor XY couplings; see Fig.~\ref{fig:sketch} and formulae below. We refer to this construction as the Anderson Quantum Sun (AQS). Our results, summarized in the phase diagrams of Fig.~\ref{fig:phase_diagram} and Tab.~\ref{tab:phases}, can be outlined as follows:
\vskip 0.1cm
\noindent (i)
For strong disorder, weak interactions with the ergodic bath turn Anderson localization into a standard MBL insulator, and sufficiently strong interactions are required to thermalize the system, as in the original QS model.

\noindent (ii) Surprisingly, for smaller disorder, the coupling of the Anderson chain to the ergodic bath leads to an unconventional, multistage crossover toward thermalization.

(a) For weak interaction, the spectrum adheres to Poisson statistics; however, the eigenstates reveal sub-volume entanglement accompanied by unusually strong system-wide correlations dominated by rare events. This behavior contrasts sharply with a genuine many-body localized (MBL) phase, in which one would expect area-law entanglement and strictly short-range correlations.

(b) Upon increasing the coupling, before reaching the conventional ergodic thermal phase for strong enough interactions, the system enters a second anomalous regime: fully thermal (volume-law entanglement) in terms of eigenstates, but with intermediate spectral statistics.
Such behavior, typical of mixed phase space dynamics~\cite{Haakebook,Serbyn16,Sierant19b}, is reminiscent of features seen in integrable and delocalized quadratic models~\cite{Swietek24}, and resembles the unconventional phase separations observed in zero-dimensional Rosenzweig-Porter and power-law random matrix ensembles~\cite{Kravtsov15, Nosov19, DeTomasi20}. However, unlike these toy models which lack spatial structure, the AQS explicitly captures the distance-dependent resonances that drive thermalization in a genuine real-space geometry.

These observations support earlier theoretical predictions of an intermediate sub-volume law regime that separates the MBL and volume-law phases~\cite{Grover14,Monthus16}. They also shed new light on recent numerical studies reporting the existence of an intermediate regime between MBL and ETH~\cite{Dumitrescu19,Weiner19,Evers23,Colbois24,Colbois24a,Tarzia2024,Laflorencie25}.

\paragraph{The U(1) symmetric quantum sun model.}
We briefly summarize the construction of the U(1) symmetric QS model~\cite{Pawlik24}, whose Hamiltonian is given by
\begin{equation}
    {\cal{H}}_{\rm QS}={\mathsf{R}}_{\rm QS}+\sum\limits_{j=1}^{L}\alpha^{u_{j}}({S}^{x}_{n_{j}} {S}^{x}_{j}+{S}^{y}_{n_{j}} {S}^{y}_{j})+\sum\limits_{j=1}^L h_j{S}^z_j.
    \label{eq:qs_hamiltonian}
\end{equation}
This spin-${1/2}$ model consists of two subsystems: $N=3$ spins internally described by a random ${2^N\times  2^N}$ matrix ${\mathsf{R}}_{\rm QS}$ from the Gaussian Orthogonal Ensemble (GOE) acting as a thermal bath for the rest of the system. Here ${\mathsf{R}}_{\rm QS}=\frac{\beta}{2}(B+B^T)$, with fixed $\beta=0.3$ and matrix elements $B_{ij}$ given by i.i.d. standard Gaussian distributions. Furthermore, all matrix elements $({\mathsf{R}}_{\rm QS})_{ij}$ that spoil commutation with the U(1) symmetry generator ${\hat{S}^z_{\mathrm{tot}}=\sum_{j=1-N}^{L}\hat{S}_j^z}$ are fixed as $0$. The second component of the model is the localized chain of $L\gg N$ spins that do not interact with each other. A random magnetic field ${h_j\in[1-W,1+W]}$ acts at each spin outside the sun, where the disorder strength is fixed as $W=1/2$. The interaction between both parts exponentially decreases with the distance $u_j$ between the localized spin and the randomly chosen spin of the sun $n_j$ as $V_{\rm AQS}=\alpha^{u_j}=\exp(-u_j/\Lambda_\alpha)$. Moreover, positional disorder $u_j\in[j-\zeta,j+\zeta]$ with $\zeta=0.2$ is introduced. Its role is purely numerical, it helps to reduce the number of disorder realizations required for convergence of the results.

\paragraph{The AQS model.}
We aim to generalize the U(1) symmetric QS model~\cite{Pawlik24} to the AQS case, where spins in the chain are now coupled to form an Anderson chain, with the Hamiltonian:
\begin{equation}
    {\cal{H}}_{\rm AQS}={\cal{H}}_{\rm QS}+J\sum_{j=1}^{L-1}({S}^{x}_{j} {S}^{x}_{j+1}+{S}^{y}_{j} {S}^{y}_{j+1}).
    \label{eq:model_hamiltonian}
\end{equation}

Note that, if coupling to the sun is neglected, the random magnetic field ${h_j\in[1-W,1+W]}$ leads to Anderson localization of the chain for any finite $W$, with an average many-body localization length~\footnote{The phenomenological expression Eq.~\eqref{eq:AL} yields an average localization length, averaged over the density of single-particle states, see Ref.~\cite{Colbois23}.} given by

\begin{equation}
    \xi_{\rm AL}={1}/{\ln\left[1+\left(\frac{W}{W_0 J} \right)^2\right]}
    \label{eq:AL}
\end{equation}
with $W_0\sim 1.2$~\footnote{As noted in Ref.~\cite{Colbois23},  $W_0$ slightly varies with $W/J$, from $1.13$ at strong disorder to $1.22$ at weak disorder.}. The interaction between the sun spins $n_j$ and the external spins $j$ decays with the distance $u_j\approx j$ exponentially $V_{\rm AQS}=\alpha^{u_j}=\exp(-u_j/\Lambda_\alpha)$. This defines the length scale associated with the interaction

\begin{equation}
    \Lambda_\alpha=1/|\ln \alpha|.
    \label{eq:Lambda}
\end{equation}

Even though the size of the sun, $N$, is fixed, the system can exhibit an ergodicity-breaking transition, as found in the original QS model ($J=0$) for $\alpha^c\sim 0.75$~\cite{Suntajs22,Pawlik24}, in good agreement with the prediction of the avalanche scenario $1/\sqrt{2}$~\cite{DeRoeck17,Luitz17}.

The above AQS model is expected to exhibit richer behavior upon varying the free parameters $\alpha$ and $J$, keeping the random-field width fixed at $W = 0.5$. Indeed, it involves a more realistic interplay between noninteracting localization, controlled by $\xi_{\rm AL}$, and interaction with an ergodic bath governed by $\Lambda_\alpha$, as we discuss now in more detail, based on intensive numerics.

\begin{table}[b!]
    \centering
    \renewcommand{\arraystretch}{1.4}
    \begin{tblr}{
            colspec={c|c|c|c|c},
            colsep=2pt,
        }
        Regime                 & Spectrum                   & ${\rm EE}\propto L^b$    & Correlations           & PE                           \\ \hline
        \SetCell[r=2]{c} {[A]} & \SetCell[r=3]{c} {Poisson} & \SetCell[r=2]{c} {$b=0$} & Short range            & \SetCell[r=3]{c} {$0<D_q<1$} \\
                               &                            &                          & No rare events         &                              \\
        \cline{1-1} \cline{3-4}
        [B]                    &                            & $0<b<1$                  & Rare events            &                              \\
        \cline{1-1}\cline[dashed]{2-5}
        [C]                    & Intermediate               & \SetCell[r=2]{c} {$b=1$} & \SetCell[r=2]{c} {ETH} & \SetCell[r=2]{c} {$D_q=1$}   \\ \cline{1-1}\cline[dashed]{2-2}
        [D]                    & GOE                        &                          &                        &                              \\
    \end{tblr}
    \caption{Properties of the different regimes of the AQS model, dashed lines corresponding to phase transitions, and labels of regimes~[A-D] are the same as in Fig.~\ref{fig:phase_diagram}.}
    \label{tab:phases}
\end{table}

\paragraph{Numerical methods.}

We perform statistical analysis of both eigenvalues and eigenvectors. In the former case, we analyze the gap ratios ${r_i=\min({s_i,s_{i+1}})/\max({s_i,s_{i+1}})}$~\cite{Oganesyan07}, where ${s_i=E_{i+1}-E_i}$ are the energy gaps. Random matrix theory predicts~\cite{Atas13, Giraud22} different distributions of gap ratios for integrable and ergodic systems, with the average gap ratio, $\langle r \rangle\approx 0.39$ for the former (corresponding to the Poisson ensemble) and $\langle r \rangle \approx 0.53$ for generalized time-reversal invariant systems~\cite{Haakebook} corresponding to the GOE.

Regarding eigenvectors, we analyze the von Neumann entanglement entropy (EE), expressed as $S(\ket{\psi})=-\Tr_{A}\rho_A\ln\rho_A$ under bipartition of the system into $A, B$ parts, where $\rho_A = \Tr_B |\psi\rangle\langle\psi|$ is a reduced density matrix for $A$ and $\Tr_B$ is a partial trace over $B$. For the $A$ part, we take the GOE sun combined with $\lfloor (N+L)/2\rfloor -N$ closest (with the ordering of spins dictated by the $u_j$) Anderson spins. Then $B$ is chosen as the remaining part of the system. The EE is averaged over the states corresponding to the selected energy window. As entropy is an extensive quantity, to compare different system sizes, we introduce normalization by Page entropy $S_P$~\cite{Page93, Bianchi22}, entropy of a Haar random vector from the Hilbert space, with the ratio $S/S_P$ going from close to $0$ value to $1$ in the limiting cases of localized and maximally extended states, respectively.

\begin{figure}[b!]
    \centering
    \includegraphics[width=\linewidth]{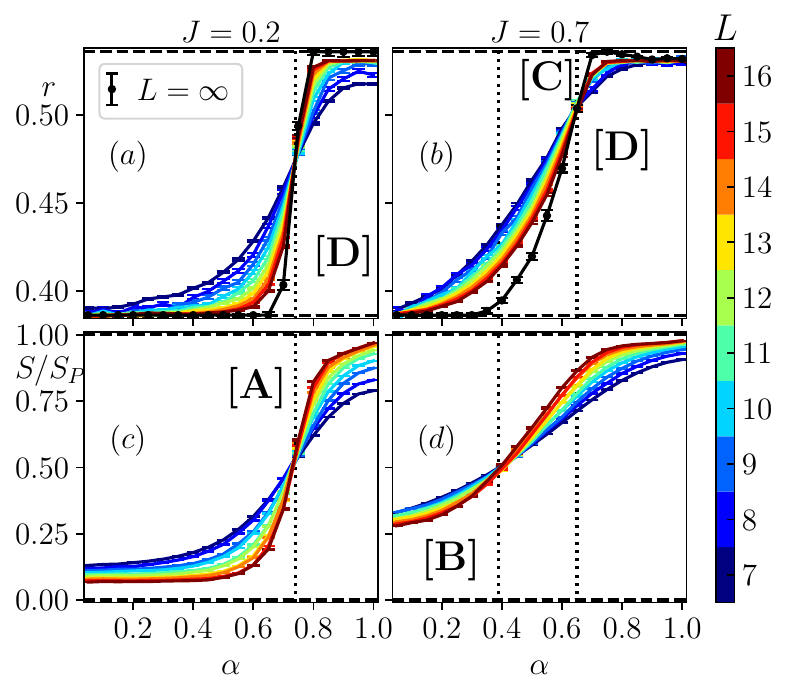}
    \caption{High energy indicators of ergodicity-breaking, as a function of parameter $\alpha$, for a fixed value of $J$. (a, b): gap ratio $r$, black circles show extrapolations to the thermodynamic limit $L=\infty$. (c, d): rescaled half-chain entanglement entropy of eigenstates $S/S_P$. Dashed lines indicate the prediction of the GOE and Poisson ensembles, while dotted lines indicate transitions. Different colors represent different system sizes - see color bar. Regimes are labeled [A-D] in accordance with Fig.~\ref{fig:phase_diagram}.}
    \label{fig:vertical_scan}
\end{figure}

\paragraph{Vertical scans.}
The results for both quantities are presented in Fig.~\ref{fig:vertical_scan} for two representative vertical scans across the $J-\alpha$ phase diagram (Fig.~\ref{fig:phase_diagram}). For small $J=0.2$ (left column), corresponding to short $\xi_{\rm AL}$, the results strongly resemble the qualitative behavior of the U(1) QS at $J=0$, with an ergodicity breaking transition observed around $\alpha^{\rm c}_r,\alpha_S^{\rm c} \approx 0.75$ for both quantities $r$ and $S$. This remains true for small enough hopping $J\leq 0.3$, corresponding to a strong disorder in the Anderson chain, i.e., $W/J \gtrsim 1.7$.

However, unexpectedly, for larger values of $J$ (i.e., weaker disorder), for instance $J=0.7$ as shown in Fig.~\ref{fig:vertical_scan} (right panels), the transition points do not coincide. The entanglement transition toward ergodic behavior occurs at a significantly smaller $\alpha^{\rm c}_S$ than the gap ratio transition $\alpha_r^{\rm c}$. Moreover, the latter transition occurs much closer to the GOE value, suggesting significant differences in the intermediate spectral statistics around the transition.

\begin{figure}[ht]
    \includegraphics[width=\linewidth]{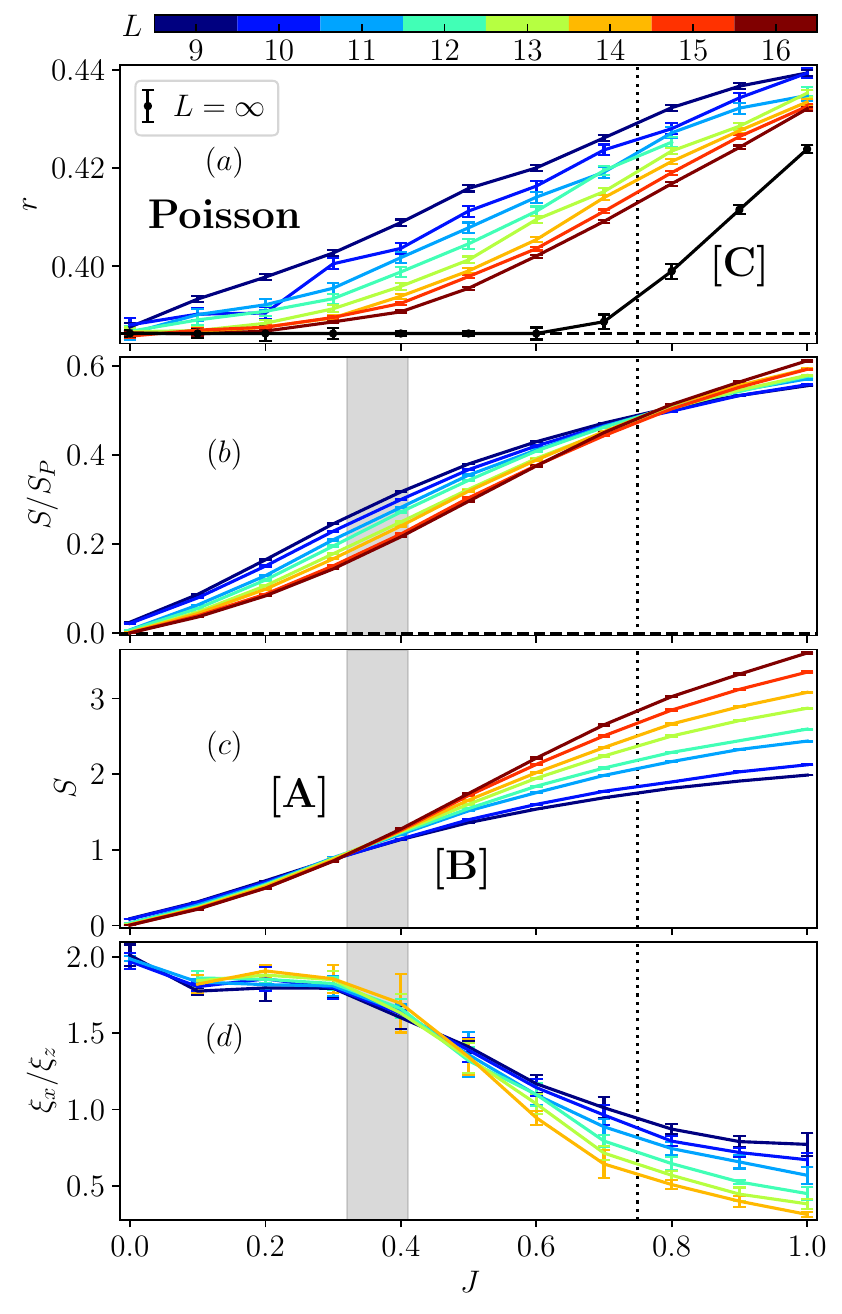}
    \caption{Horizontal scan of the phase diagram at $\alpha=0.35$. (a): Gap ratio $r$. (b): Rescaled entropy $S/S_P$. (c): Entanglement entropy $S$. (d): Correlation length ratio $\xi_x/\xi_z$ of $C_{1,L}$, for different choices of the fitting window of system sizes $[L-2,L+2]$, with $L$ indicated by the color bar. In panels (a, b), dashed horizontal lines indicate the Poisson limit, and the black solid curve in (a) represents the extrapolation to the thermodynamic limit. The dotted line indicates~[B]-[C] transition based on $S/S_P$, while the gray band corresponds to the onset of power law $S$-dependence in~[A]-[B] transition. Data points for $L=12$–$14$ in (d) at $J=0$ are omitted, as they contain events with $C_{1,L}$ below numerical precision.}
    \label{fig:horizontal_scan}
\end{figure}

\paragraph{Horizontal scan.}
To gain additional insight into the discrepancy between spectral and eigenstate behaviors, we follow a horizontal cut in the phase diagram, by varying the hopping $J$ at fixed $\alpha=0.35$, see Fig.~\ref{fig:horizontal_scan}. By extrapolating the gap ratio $r$ in Fig.~\ref{fig:horizontal_scan}(a) to the thermodynamic limit (details in~\cite{suppl}), we see that Poisson statistics is present only in the region of nonergodic EE, i.e. if $S/S_P$ decreases with the system size $L$, see Fig.~\ref{fig:horizontal_scan}(b). For larger $J$, EE shows an ergodic behavior, with $S/S_P\to 1$, while the gap ratio $r$ approaches an intermediate value between Poisson and GOE statistics. This suggests that the whole region between the critical points $\alpha^{\rm c}_S$ and $\alpha^{\rm c}_r$ (corresponding to the transition in $S/S_P$ and $r$, respectively) will share this behavior, which is confirmed by the phase diagram in Fig.~\ref{fig:phase_diagram}(a). A closer look at the system size dependence $S(L)$ in Fig.~\ref{fig:horizontal_scan}(c) reveals that the region of nonergodic EE is divided into two distinct regimes, one showing genuine area-law entanglement~[A], and an unusual extended one~[B] where a sub-volume EE is observed.

To better understand these intriguing findings, we also analyze the connected correlation function of the eigenstates, following Ref.~\cite{Colbois24}:
\begin{equation}
    C_{ij}^{\gamma\gamma}=\left|\langle S_i^\gamma S_j^\gamma\rangle-\langle S_i^\gamma \rangle\langle S_j^\gamma\rangle\right|,
\end{equation}
with $\gamma=x,z$. We choose $i=0,1$ ($0$ representing a random spin in the sun) and $j=\lfloor L/2\rfloor,L$, giving $4$ types of correlation with extensive $|i-j|$. The typical value of the connected correlation for localized states is expected to decay exponentially with distance $|i-j|$, whereas in the ergodic phase the randomness of the state determines the scaling: exponential with the inverse of the Hilbert space dimension for $\gamma=x$, and, due to U(1) symmetry, power-law $\sim L^{-1}$ for $\gamma=z$~\cite{Pal10,Colbois24}.

We define the typical correlation lengths $\xi_\gamma$ of the decays
\begin{equation}
    \overline{\ln C_{ij}^{\gamma\gamma}}=-\frac{|i-j|}{\xi_\gamma}+\mathcal{O}(1),
\end{equation}
where $\overline{(\cdot)}$ denotes disorder and eigenstates averaging. If $C_{ij}^{zz}$ decays algebraically, fitting to the above form fails, and the extracted values of $\xi_z$ will effectively grow with the system size. Taking $\xi_x$ as the reference value, we can use $\xi_x/\xi_z$ to detect the signatures of ergodic instabilities~\cite{Colbois24,Colbois24a} (corresponding to the onset of systemwide resonances), as shown in Fig.~\ref{fig:horizontal_scan}(d) for $C_{1,L}$. With increasing system size $L$, it would approach $\xi_x/\xi_z\to2$ for an Anderson model or a strongly disordered XXZ Heisenberg chain~\cite{Colbois24}, which is also the behavior of the AQS for small nonzero $J$ and small $\alpha$.

Remarkably, the ergodic instabilities revealed by decreasing $\xi_x/\xi_z$ with $L$, show up in the same regime~[B] where sub-volume EE is observed, see Fig.~\ref{fig:horizontal_scan}(c), suggesting a link between the two. These results are supported by all types of correlations considered, see~\cite{suppl}. Interestingly, even at $J=0$, the QS model retains a correlation length ratio $\xi_x/\xi_z \approx 2$.

\paragraph{Multifractal analysis.}

\begin{figure}[b!]
    \centering
    \includegraphics[width=\linewidth]{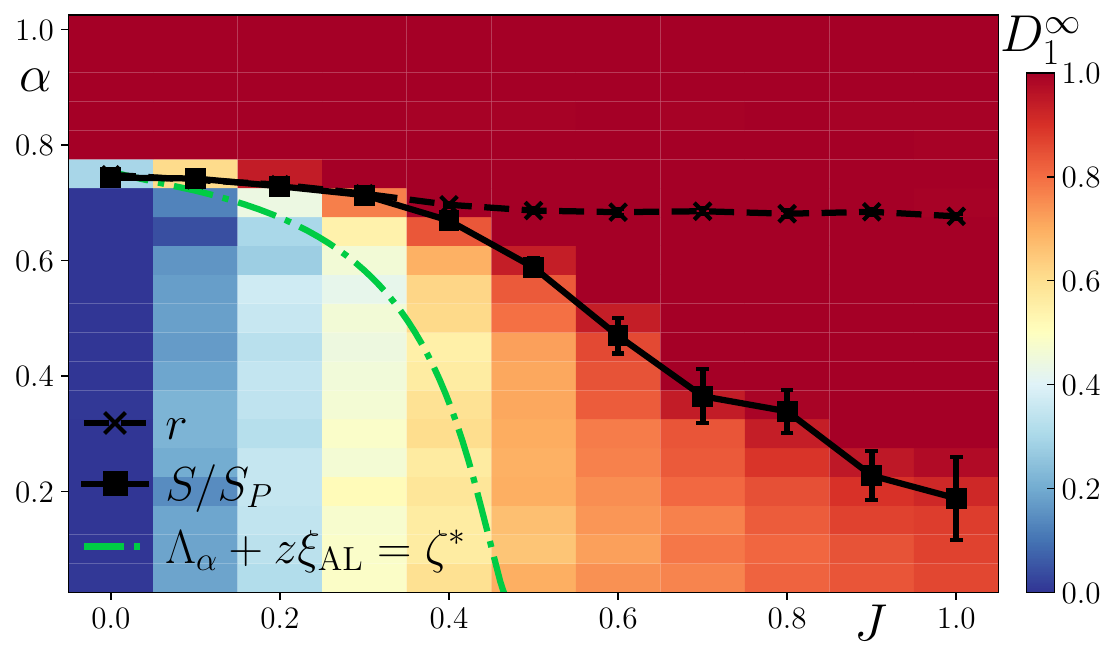}
    \caption{Multifractal dimension $D_1$ of eigenstates in the AQS model. For explanations of the lines, see Fig.~\ref{fig:phase_diagram} of the main text.}
    \label{fig:phase_diagram_multifractal}
\end{figure}

To complement the study of eigenstates $\ket{\psi}$, we consider participation entropy (PE) $P_q$~\cite{Mace19b}

\begin{equation}
    \ket{\psi}=\sum_{\alpha=1}^\mathcal{D}\psi_\alpha\ket{\alpha},\quad P_q=\frac{1}{1-q}\ln\left[\sum_{\alpha=1}^\mathcal{D}|\psi_\alpha|^{2q}\right],
\end{equation}
where $\mathcal{D}$ is the Hilbert space dimension, and $\ket{\alpha}$ denotes a spin product state. In particular, $P_1=-\sum_\alpha|\psi_\alpha|^2\ln|\psi_\alpha|^2$. This quantity captures how the state spreads over the computational basis, with $P_q = \rm const$ for a Fock-space localized system and $P_q = \ln \mathcal{D}$ for an ergodic one. Intermediate multifractal behavior is described by
\begin{equation}
    P_q=D_q\ln\mathcal{D}+b_q.
\end{equation}
Scaling analysis of $P_q$, following a procedure analogous to that used for entanglement entropy in Fig.~\ref{fig:phase_diagram}(b), gives Fig.~\ref{fig:phase_diagram_multifractal}.

The region of volume-law entanglement entropy aligns perfectly with $D_1=1$, suggesting fully ergodic eigenstates in that regime. By contrast, the area-law~[A] and sub-volume~[B] regimes on the phase diagram (Fig.~\ref{fig:phase_diagram}) are characterized by the fractal dimension inherited from the Anderson chain, corresponding to $\alpha=0$. The Fock-space localization is observed only in the QS model ($J=0$), consistent with Ref.~\cite{Suntajs24s}.

\paragraph{Phase diagram.}
The two panels of Fig.~\ref{fig:phase_diagram} show the global phase diagrams from two different viewpoints: (a) spectral statistics and (b) eigenstate entanglement. In Fig.~\ref{fig:phase_diagram}(a), we can see that for $J\geq 0.4$, a gap opens between the GOE and Poisson regimes, beyond statistical uncertainties. This indicates a new regime with volume-law entanglement and intermediate statistics~[C]. For smaller values of $J$, this extended unconventional regime shrinks and seems to merge into a critical line that ends at the QS critical point, which also exhibits intermediate statistics~\cite{Pawlik24,Suntajs24s}. It seems that introducing a sufficiently strong coupling $J$ elevates the single QS critical point to a whole phase, intervening between GOE and Poisson statistics.

We now turn to the unconventional entanglement properties, illustrated by the sub-volume region~[B] in Fig.~\ref{fig:horizontal_scan}(c). To probe the intermediate EE scaling, we perform a local fit $S\propto L^{b(L)}$ for a narrow window $[L-2,L+2]$ of system sizes and extrapolate the $b(L)$ results to $L\to\infty$, which is depicted in Fig.~\ref{fig:phase_diagram}(b). The region showing a genuine volume-law $b=1$ is in excellent agreement with the extrapolated eigenstate phase transition (solid curve). On the other hand, below this transition, the phase with Poisson statistics splits into two regimes: an area-law region~[A] with $b\approx0$, and a sub-volume regime~[B] where $0<b<1$. It is worth noting that this region~[B] also hosts ergodic instabilities and rare events, observed in the system-wide correlations, see~\cite{suppl}.

To better understand the crossover between~[A] and~[B] inside the Poisson regime,  we consider the two characteristic length scales of the model: the average Anderson localization length $\xi_{\rm AL}$ [Eq.~\eqref{eq:AL}] and the length scale $\Lambda_\alpha$ [Eq.~\eqref{eq:Lambda}] associated with the interaction with the sun. They clearly play different roles in the AQS model, but both control how the system gets delocalized. Increasing any of these scales clearly enhances delocalization, and if an avalanche instability exists, it could follow a generalized avalanche threshold of the form

\begin{equation}
    \Lambda_\alpha+z\xi_{\rm AL}=\zeta^*,
\end{equation}
where $z$ and $\zeta^*$ are constants depending on the microscopic details of the model. We observe that $z\approx 2$ and $\zeta^*\approx 3.7$ give a good description of the crossover between~[A] and~[B], see the green dotted-dashed line in Fig.~\ref{fig:phase_diagram}(b).

At this stage, it is important to note that the presence of the sub-volume region in the AQS model is in stark contrast to the XXZ model, where area-law entanglement entropy and Poisson level statistics always appear together~\cite{Colbois24}. The properties of all regimes are summarized in Tab.~\ref{tab:phases}, together with the participation entropy.

\paragraph{Discussion.}

Our study reveals a surprising breakdown of the commonly observed correspondence between eigenstate localization and spectral statistics in many-body interacting systems. While paradigmatic models of MBL typically exhibit consistent signatures of localization in both eigenstates and level statistics, we conversely find for the interacting AQS the emergence of a volume-law entangled phase that retains intermediate level statistics in the thermodynamic limit. We also report the existence of an extended intermediate sub-volume regime associated with anomalously strong system-wide correlations that coexist with Poisson spectral statistics.

We note that apparent mismatches between spectral and eigenstate properties have been the subject of intense debate in non-interacting Anderson models on random regular graphs (RRGs) and tree-like structures, where they are often attributed to massive finite-size effects \cite{Tikhonov16, GarciaMata17,Biroli18, Tikhonov19}. However, unlike the finite-size crossovers in boundary-less RRGs---where the crossing point of spectral statistics consistently drifts towards the Poisson value as system size increases---the crossing point of the gap ratio $r$ in the AQS model stabilizes at a size-independent value. Furthermore, the intermediate sub-volume entanglement regime [B] is explicitly driven by long-range resonances that show no numerical indication of vanishing at larger scales. This distinct critical behavior strongly suggests that the intermediate regimes observed here represent robust physical mechanisms rather than transient finite-size artifacts.

Moreover, our results for the AQS model enrich our understanding of how Anderson localization can break down due to interactions. At small $J/W$, MBL is continuously connected to the Anderson insulator, similar to the behavior found in the disordered XXZ chain at large disorder~\cite{Colbois24}. However, at larger $J/W$, the situation differs from XXZ. Here weak interactions $\alpha$ do not change the spectral statistics but lead to delocalized sub-volume eigenstates, between the area-law and volume-law regimes. This may be a concrete realization of the intermediate region proposed by Grover in Ref.~\cite{Grover14}. Intriguingly, this regime also displays system-wide instabilities in correlation functions, reminiscent of the recently studied intermediate region(s) of MBL in the XXZ model~\cite{Colbois24,Colbois24a,Tarzia2024,Laflorencie25}.

An experimental realization of QS or AQS models remains an open challenge. One of the possibilities would be to use the flexibility of ion chains (for a review see e.g. \cite{Blatt12}). While most of the effort with ion chains is oriented towards quantum computing \cite{Cirac95,Cirac00,Bruzewicz19}, they offer an ideal medium as a quantum simulator with the ability to shape various interactions and tunnelings.

In summary, our results highlight the AQS as a promising model for studying the interplay of localization, ergodicity, and spectral statistics. They open up several avenues for future research, including a rigorous two-parameter scaling analysis to formally map the renormalization-group flow of this transition~\cite{Altshuler25, Niedda25,Swietek25b}, as well as the stability of the sub-volume phase, its potential connection to MBL studies, and its accessibility in current experimental platforms, and the critical open challenge of identifying the broader universal class of physical models that may harbor these unconventional phase separations.

\paragraph{Data Availability.}
The data required to reproduce the figures in this study are openly available in the RODBUK repository~\cite{dataset_ref}.

\acknowledgements
We thank A.~Chandran, P.~Crowley and P.R.N.~Falc\~ao for interesting discussions. We gratefully acknowledge the Polish high-performance computing infrastructure PLGrid (HPC Center: ACK Cyfronet AGH) for providing computer facilities and support within the computational grant no. PLG/2025/018457. This research has been funded by the National Science Centre (Poland) under project 2021/43/I/ST3/01142 -- OPUS call within the WEAVE programme. The study was funded by the “Research support module” as part of the “Excellence Initiative – Research University” program at the Jagiellonian University in Kraków. This study has been (partially) supported through the grant NanoX n° ANR-17-EURE-0009 in the framework of the ‘ Programme des Investissements d'Avenir’.
NL is supported by the ANR research grant ManyBodyNet No.
ANR-24-CE30-5851, and also benefited from the support of the Fondation Simone et Cino Del Duca.

\input{arxiv.bbl}

\newcommand{\snum}{S}

\renewcommand{\theequation}{\snum.\arabic{equation}}
\renewcommand{\thefigure}{\snum.\arabic{figure}}

\setcounter{equation}{0}
\setcounter{figure}{0}


\onecolumngrid

\appendix

\section{Supplementary material for Unconventional Thermalization of a Localized Chain Interacting with an Ergodic Bath}

\section{Numerical details}

Our analysis is restricted to the largest symmetry sector of the model, namely $S^z_{\rm tot}=0$ for even $N+L$, and $S^z_{\rm tot}=-1/2$ otherwise. For small system sizes, $N+L\leq 13$, we perform full exact diagonalization of the AQS model, Eq.~\eqref{eq:model_hamiltonian} in the main text, while for larger systems $13<N+L\leq 20$ the POLFED algorithm~\cite{Sierant20p} is used. We take into consideration the eigenenergies in the largest density of states window $\varepsilon\in [0.5-\varepsilon_0, 0.5+\varepsilon_0]$ (but not less than $30$ energies), with ${ \varepsilon=\frac{E-E_{\min}}{E_{\max}-E_{\min}} }\in[0,1]$ and $\varepsilon_0=0.005$. Those are disorder averaged with $10^4$ realizations for $L=7$, with $3\cdot 10^3$ for $7<L<16$  and $10^3$ for larger systems.

A more quantitative study of phase transitions is possible using a standard approach~\cite{Sierant20p,Sierant23f} of estimating the critical point at finite size $\alpha^{\rm c}_{r,L}$, $\alpha^{\rm c}_{S,L}$. Consistent scaling of this value with system size $L$ allows to extrapolate the results to $L\to\infty$, obtaining the phase boundaries $\alpha_{r,\infty}^{\rm c}$, $\alpha_{S,\infty}^{\rm c}$, which is described in detail in Sec.~\textbf{Extraction and scaling of the critical point}.

\section{Ergodic instability in correlations: correlation length ratios}

\begin{figure}[ht]
    \centering
    \includegraphics[width=0.7\linewidth]{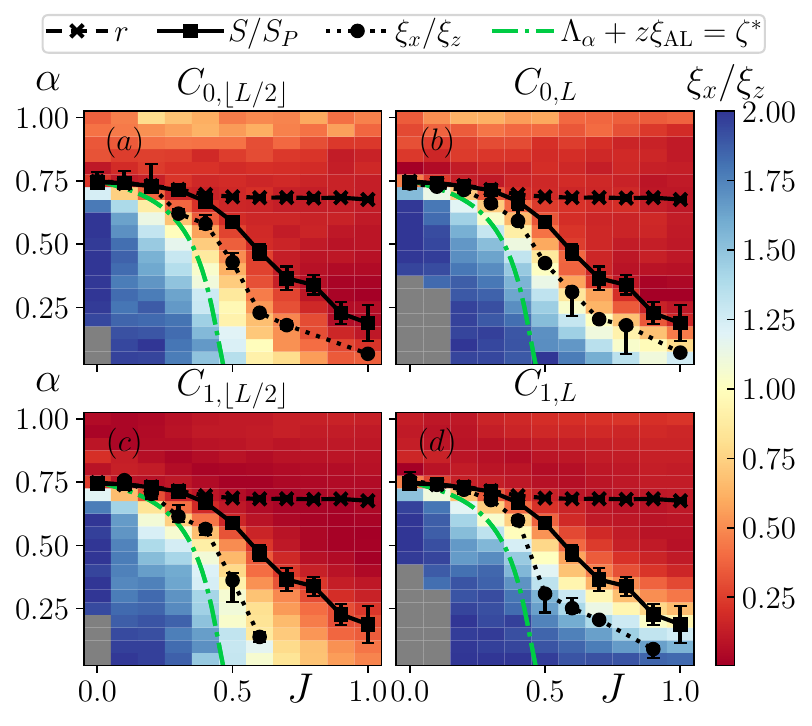}
    \caption{Phase diagram of AQS model in terms of correlation length ratios $\xi_x/\xi_z$, type of the correlation indicated above each panel. Spectral ($r$) and entanglement entropy ($S/S_P$) transitions are the same as in Fig.~\ref{fig:phase_diagram}. The $\xi_x/\xi_z$ dotted line marks the onset of ergodic instability, with the smallest $\alpha$ that has a noticeable decrease of this ratio with system size $L$. Color map corresponds to a fit over the system sizes $L\in[12,16]$. Points on the phase diagram where correlations in some states fell below numerical precision are excluded and shown in gray.}
    \label{fig:phase_diagram_xi}
\end{figure}

The regime~[B] of sub-volume entanglement growth (see Fig.~\ref{fig:phase_diagram} in the main text for the labeling of regions) exhibits ergodic characteristics in correlation functions, evidenced by a decreasing correlation length ratio $\xi_x/\xi_z$ with increasing system size $L$ in the fit $[L-2,L+2]$, as shown in Fig.~\ref{fig:horizontal_scan}(d) for the correlation $C_{1,L}$. This trend is robust for any correlation over an extensive distance $|i-j|$; in Fig.~\ref{fig:phase_diagram_xi}, we demonstrate it for $C_{i,j}$, $i = 0,1$ (with $0$ representing a random spin from the sun) and $j = \lfloor L/2 \rfloor, L$. We note that the correlations with $j=\lfloor L/2\rfloor$ behave slightly differently for even and odd system sizes $L$, thus the fit to extract the correlation lengths $\xi_x,\xi_z$ is performed for $\{L-2,L,L+2\}$, without mixing system sizes of different parity.
In addition to marking the spectral (black dashed line) and eigenstate (black solid line) transitions, we also indicate the onset of ergodic instability with a black dotted line. For each $J$, this line corresponds to the smallest value of $\alpha$ at which a statistically significant decrease in $\xi_x/\xi_z$ is observed.

\subsection{Rare events of large system-wide correlations}

\begin{figure}[ht]
    \centering
    \includegraphics[width=0.7\linewidth]{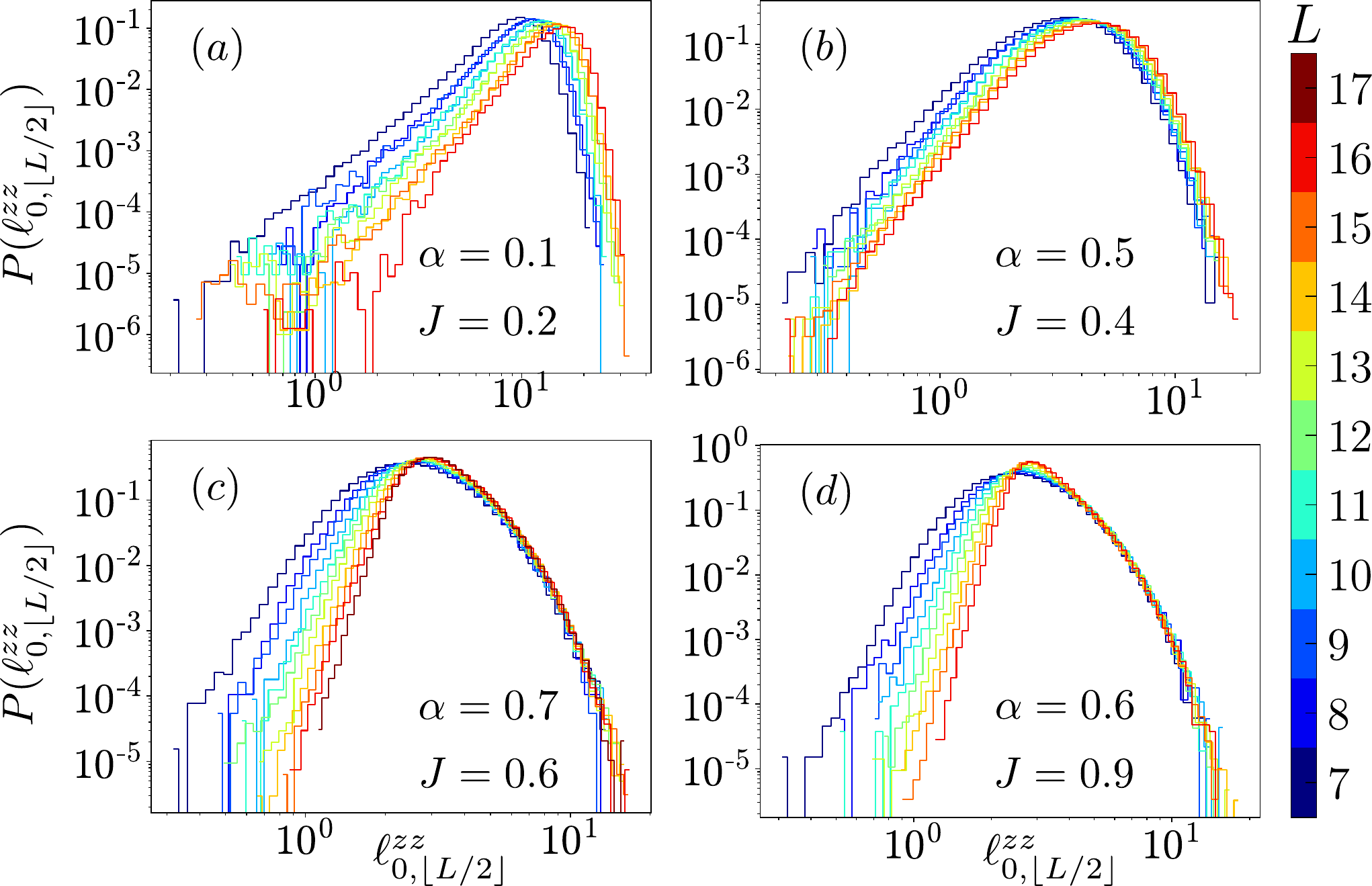}
    \caption{Histograms of rescaled correlation function $\ell^{zz}_{0,\lfloor L/2 \rfloor}$ for eigenstates of the AQS model. Panels (a-d) correspond to representative $\alpha,J$ points from regime~[A-D], respectively. System size is indicated by the color map.}
    \label{fig:corr_hist}
\end{figure}

To investigate rare events involving system-wide correlations, we introduce a rescaled correlation quantity, $\ell^{\gamma\gamma}_{ij}$
\begin{equation}
    \ell^{\gamma\gamma}_{ij}=-\ln |4C^{\gamma\gamma}_{ij}|,
\end{equation}
with $\gamma=x,z$. This transformation expands the region corresponding to large correlations, $|C^{\gamma\gamma}_{ij}|\approx 1/4$, mapping them to small values of rescaled correlation, while weak correlations yield large $\ell^{\gamma\gamma}_{ij}$. The numerical histograms for $\ell^{zz}_{0,\lfloor L/2 \rfloor}$ are shown in Fig.~\ref{fig:corr_hist}, where all regimes~[A-D] from the phase diagram of the AQS model (Fig.~\ref{fig:phase_diagram}) are analyzed.

In the strongly localized regime~[A] (Fig.~\ref{fig:corr_hist}(a)), the number of large correlation events exponentially decreases with the system size $L$, similarly to the behavior in the deep MBL regime of the XXZ model~\cite{Colbois24a}. In contrast, the region~[B] (Fig.~\ref{fig:corr_hist}(b)), which exhibits ergodic instability in correlation functions, shows a power-law tail at small values of $\ell$, involving a significant fraction of all eigenstates. This feature resembles the recently proposed intermediate region between the MBL and ergodic phases of the XXZ model~\cite{Colbois24a,Laflorencie25}. Meanwhile, regions~[C] and~[D] (Fig.~\ref{fig:corr_hist}(c,~d)) show a rapid suppression of such large-correlation events with system size, consistent with expectations for random eigenstates obeying ETH.

Based on the observed ergodic instabilities and rare events, two possible scenarios arise for the sub-volume region~[B]: thermal avalanches may ultimately drive it towards ergodicity with volume-law entanglement growth, or, alternatively, these effects may remain insufficient, consistent with our EE results, leaving the region nonergodic.

\subsection{Probability distributions of $r$ and $S$}
\begin{figure}[ht]
    \centering
    \includegraphics[width=0.7\linewidth]{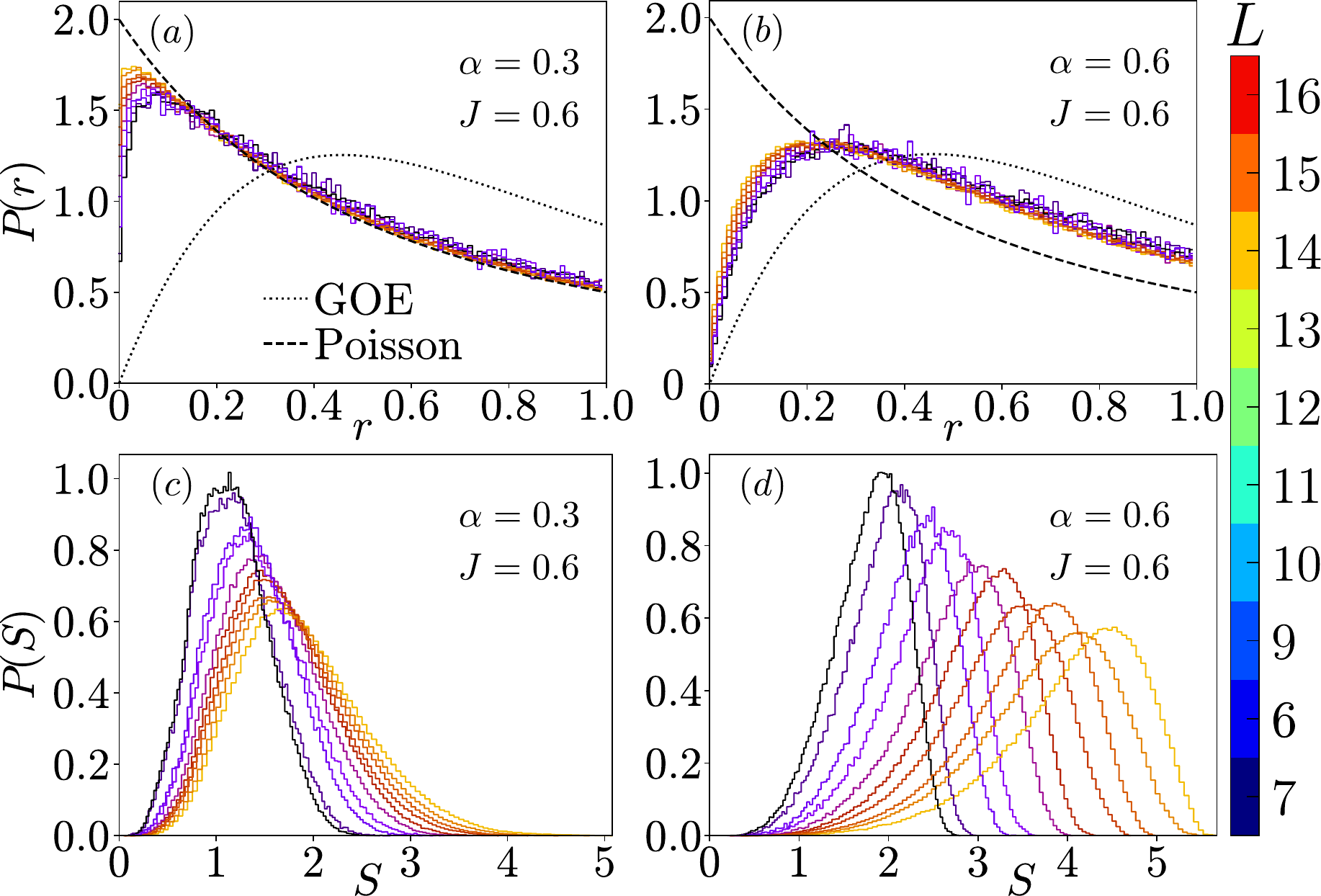}
    \caption{Histograms of $r$ (a,b) and $S$ for the AQS model. Panels (a,c) correspond to representative $\alpha,J$ point from regime~[B], while (b,d) correspond to phase~[C]. System size is indicated by the color map.}
    \label{fig:r_S_hist}
\end{figure}

We present distributions $P(r)$ and $P(S)$ for representative points of regions~[B] and~[C] in Fig.~\ref{fig:r_S_hist}. In~[B], Fig.~\ref{fig:r_S_hist}(a,~c), $P(r)$ follows  the Poisson distribution, while $P(S)$ shows power-law increase of the average value. Phase~[C], Fig.~\ref{fig:r_S_hist}(b,~d), manifests volume-law entanglement, with intermediate $P(r)$ distribution. In phases~[A] and ~[D], not shown, distributions fully agree with localized and ergodic systems, respectively.

\section{Statistical uncertainties}
In the analysis of the main text, we considered an observable $O$ averaged over an energy window, but such single disorder realization averages $O_S$ are known not to be self-averaging in some systems~\cite{TorresHerrera20,TorresHerrera20b}, which means that its variance ${\big\langle (O_S-\langle O_S\rangle)^2\big\rangle}$, where $\langle\cdot\rangle$ means average over disorder realizations, may not decrease with increasing system size. Instead, one needs to consider quantities averaged over disorder realizations $\langle O_S\rangle$, which is denoted $\langle O\rangle$ for short. As we consider uncorrelated realizations, we may compute the standard deviation of such an average as~\cite{Sierant23f}:

\begin{equation}
    \sigma_O= \frac{\big\langle (O_S-\langle O \rangle)^2\big\rangle^{1/2}}{N_{\mathrm{dis}}^{1/2}},
\end{equation}
where $N_{\mathrm{dis}}$ is the number of disorder realizations used for the calculation of $\langle O\rangle$.

The estimate of the critical point is obtained using two distinct methods, described in detail in Sec.~\textbf{Extraction and scaling of the critical point}. Due to a different character of these methods, the corresponding uncertainties are computed differently.

For the method based on crossings of numerical curves, we consider two single-parameter functions $f_1(\alpha)$ and $f_2(\alpha)$ with pointwise uncertainties $\sigma_1(\alpha)$ and $\sigma_2(\alpha)$. The function $\chi^2(\alpha)$ is defined:
\begin{equation}
    \chi^2_L(\alpha)=\frac{[f_1(\alpha)-f_2(\alpha)]^2}{\sigma_1^2(\alpha)+\sigma_2^2(\alpha)}.
\end{equation}
The zeros of the function $\chi^2(\alpha_0)$ correspond to intersection points $\alpha_0$ of the two functions. Near such a point, $\chi^2(\alpha)$ exhibits a quadratic minimum. If the curves are well separated away from the crossing (i.e., not approximately parallel), the equation $\chi^2(\alpha)=1$ has two solutions $\alpha_-<\alpha_+$. The quantity $(\alpha_+-\alpha_-)/2$ has a statistical interpretation of the uncertainty of $\alpha_0$~\cite{Cowan98}. The standard uncertainty propagation is then used to estimate the uncertainty of the function value at the crossing, $f_1(\alpha_0)=f_2(\alpha_0)$.

The second method determines the critical point by locating where a curve reaches a fixed threshold value. Here, the uncertainty is estimated via bootstrapping: we repeat the extraction procedure $1000$ times, each time by sampling data points from normal distributions with mean $f(\alpha)$ and standard deviation $\sigma(\alpha)$. The standard deviation of the resulting set of critical point estimates is taken as the uncertainty of the critical point.

\section{Extraction and scaling of the critical point}
To estimate the critical point of the model in the thermodynamic limit, a measure of the critical point for finite system sizes $L$ must be introduced, and then these results must be extrapolated $L\to \infty$. We employ two definitions of the critical point, the first based on crossing of the finite-size curves, the other using a threshold value near GOE.

\begin{figure}[ht]
    \centering
    \includegraphics[width=0.95\linewidth]{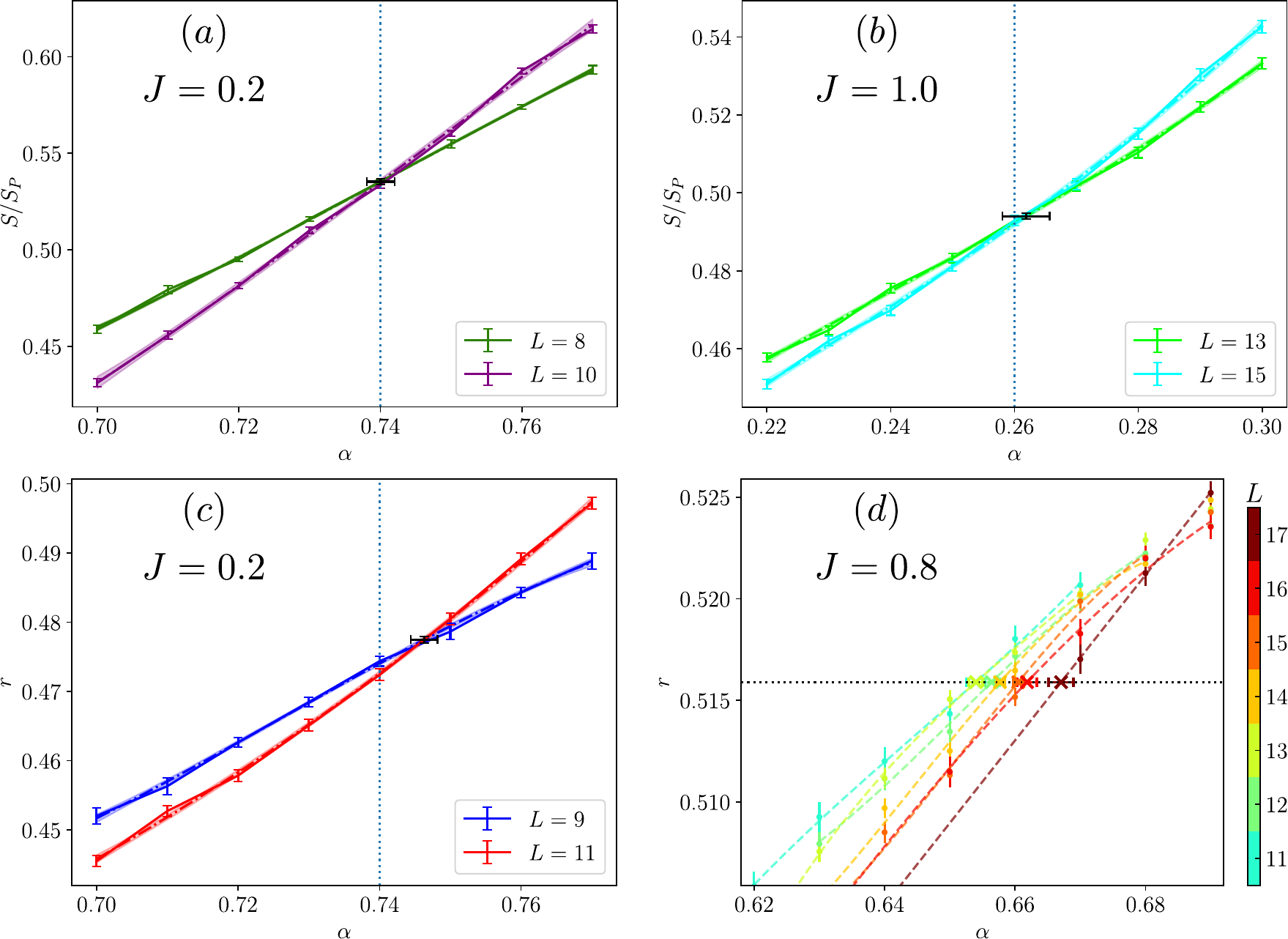}
    \caption{Extraction of the critical point. Crossing method: curves corresponding to system sizes $L-1$ and $L+1$ for (a, b): rescaled entanglement entropy $S/S_P$ and (c): gap ratio $r$. A dotted vertical line corresponds to visual estimate $\alpha_0'$, while black error bars give uncertainties for the coordinates of the crossing point. Threshold method (d): numerical curves and polynomial fits used for the extraction. Horizontal dotted line indicates $f_{\rm thresh}=0.02$, crosses show obtained values of the critical point together with uncertainty.}
    \label{fig:example_crossings}
\end{figure}

For the crossing method, let $f_1(\alpha)$ be a numerical curve for $J$, the system size $L-1$, and the given observable ($r$ or entropy), then $f_2(\alpha)$ is an analogous curve for $L+1$. Furthermore, let $\sigma_1,\sigma_2$ be the corresponding uncertainties. We can define $\alpha_L^{\rm c,cross}$ as their crossing point $\alpha_0$ with its standard deviation $(\alpha_+-\alpha_-)/2$, following the notation of the Sec.~\textbf{Statistical uncertainties}. To obtain the intersection numerically, we first visually estimate $\alpha_0'$ where it is expected, and then interpolate $f_1$ and $f_2$ with polynomials of degree $3$ in a range $[\alpha_0'-0.04,\alpha_0'+0.04]$. The small interpolation range eliminates accidental crossings. Example extractions are shown in Fig.~\ref{fig:example_crossings}.

In the threshold method, we again take numerical curves $f(\alpha)$, which correspond to either the gap ratio $r$ or the entropy $S/S_P$, and then extract $\alpha_L^{\rm c,thresh}$ such that

\begin{equation}
    f(\alpha_L^{\rm c, thresh})=f_{\rm GOE} - f_{\rm thresh}.
\end{equation}
In extraction of $\alpha_L^{\rm c,thresh}$, we fitted the numerical data $f(\alpha)$ with a polynomial of degree $3$ in a small window of $\alpha$ near the GOE value $f_{\rm GOE}$, avoiding the region where $f(\alpha)$ saturates on the value $f_{\rm GOE}$. The example of this procedure is presented in Fig.~\ref{fig:example_crossings}(d).

For the eigenstate data, $S/S_P$, the crossing method is more reliable, as the curves intersect well away from the limiting values. We note that the threshold method for relatively large $f_{\rm thresh}=0.4$ yields nearly identical estimates of the transition at the thermodynamic limit $\alpha^{\rm c}_{\infty}$. In contrast, for the spectral gap ratio $r$, the crossing method is optimal only for small coupling $J \leq 0.2$; at larger $J$, the crossings occur too close to the GOE value $r_{\rm GOE}$, making the threshold method more suitable. For $J \geq 0.3$, we adopt $f_{\rm thresh}=0.02$, although other values do not result in any qualitative changes. In applying the threshold method, we consider only system sizes $L > 10$ to mitigate finite-size effects. The critical point determined using the preferred method in each case is denoted as $\alpha_L^{\rm c}$.

\begin{figure}[ht]
    \centering
    \includegraphics[width=0.95\linewidth]{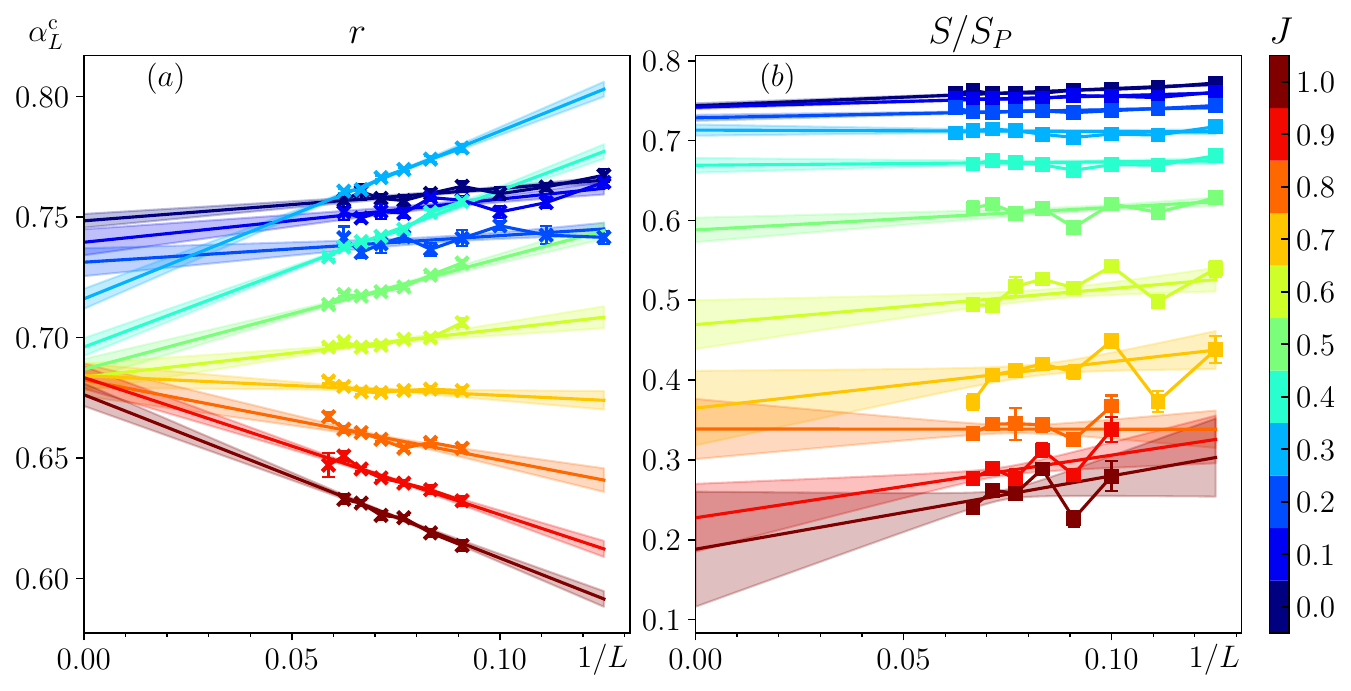}
    \caption{Extracted values of $\alpha_L^{\rm c}$ for the spectral gap ratio $r$ and for rescaled half-chain entanglement entropies $S/S_P$. Colors correspond to values of $J$. Lines with shaded error band correspond to linear fits in $1/L$. }
    \label{fig:all_crossings}
\end{figure}

The finite-size estimation of the critical point $\alpha^{\rm c}_L$ follows a system size dependence:
\begin{equation}
    \alpha^{\rm c}_L=a/L+\alpha^{\rm c}_\infty,\label{eq:extrapolation}
\end{equation}
which enables extrapolation of the critical point to the thermodynamic limit~\cite{Sierant20p,Sierant23f}, see Fig.~\ref{fig:all_crossings}. For the gap ratio $r$, the crossings $\alpha_L^{\rm c,cross}$ (Fig.~\ref{fig:all_crossings}(a) for $J\leq0.2$) follow a finite-size scaling different from $\alpha_L^{\rm c,thresh}$ (Fig.~\ref{fig:all_crossings}(a) for $J\geq0.3$), the former trying to directly estimate the critical point, while the latter serves as the upper bound instead. Nonetheless, both are well described by the scaling~\eqref{eq:extrapolation}, yielding consistent estimates of $\alpha^{\rm c}_\infty$ and resulting in its smooth dependence on $J$, which can be seen on the phase diagrams shown in this work. For the entanglement $S/S_P$, however, the crossings are consistent with~\eqref{eq:extrapolation} for all considered values of $J$, as shown in Fig.~\ref{fig:all_crossings}(b).

In all other extrapolations throughout this work, we also adopt Eq.~\eqref{eq:extrapolation}.

\section{Relation between $x$ and $z$ correlations}
Using Jordan-Wigner transformation to fermions $c_i$, correlation functions read~\cite{Colbois24, Colbois24a}:
\begin{equation}
    C_{ij}^{zz}=-\left|\langle c_i^\dagger c_j\rangle\right|^2,\quad C_{ij}^{xx}=\frac{1}{2}\langle c_i^\dagger (-1)^{\varphi_{ij}} c_j \rangle,
\end{equation}
with Jordan-Wigner phase factor $\varphi_{ij}=\sum_{l=i}^{j-1}c_l^\dagger c_l$. For sufficiently strong disorder, local densities $\langle c_i^\dagger c_i\rangle$ are close to either $0$ or $1$, so phase factor $(-1)^{ \varphi_{ij}}$ effectively reduces to $\pm1$, which leads to:
\begin{equation}
    C_{ij}^{xx}\approx \pm \frac{1}{2}\sqrt{|C_{ij}^{zz}|},
\end{equation}
which implies the relation between typical correlation lengths:
\begin{equation}
    \xi_x=2\xi_z.
    \label{eq:xi_x_vs_xi_z}
\end{equation}

At small disorder, when this mean-field argument is not applicable, one can use the exact expression for $C^{xx}_{ij}$~\cite{Henelius98,Colbois24a} in terms of $C^{zz}_{kl}$. For that, let us fix $i,j$ in $C^{xx}_{ij}$ that we want to expand and consider correlation $G_{kl}=\langle (c_k^\dagger-c_l)(c_k^\dagger+c_l)\rangle$. In the case of conserved number of particles, relevant for AQS model, it simplifies to $G_{kl}=2\langle c_k^\dagger c_l\rangle-\delta_{kl}$, where we can denote $F_{kl}=-2\langle c_k^\dagger c_l\rangle$. Using Wick's theorem, we can express $C_{ij}^{xx}$ in terms of $G_{kl}$~\cite{Henelius98,Colbois24a}, and then with $F_{kl}$ as
\begin{equation}
    C_{ij}^{xx}=\frac{(-1)^{i-j}}{4}
    \begin{vmatrix}
        F_{i,i+1}     & F_{i,i+2}     & \cdots & F_{i,j-1}     & F_{i,j}   \\
        1+F_{i+1,i+1} & F_{i+1,i+2}   & \cdots & F_{i+1,j-1}   & F_{i+1,j} \\
        F_{i+2,i+1}   & 1+F_{i+2,i+2} & \cdots & F_{i+2,j-1}   & F_{i+2,j} \\
        \vdots        & \vdots        & \ddots & \vdots        & \vdots    \\
        F_{j-1,i+1}   & F_{j-1,i+2}   & \cdots & 1+F_{j-1,j-1} & F_{j-1,j}
    \end{vmatrix}
    =\frac{(-1)^{i-j}}{4}\det(\mathrm{I}_{-1}+F),
\end{equation}
where $\mathrm{I}_{-1}$ is a matrix with $1$ only right below the main diagonal, and $F$ is a matrix of $F_{kl}$, for $k\in\{i,\ldots,j-1\}$, $l\in\{i+1,\ldots,j\}$. Using a well known identity

\begin{equation}
    \det(A+B)=\det A+\mathrm{Tr}[\mathrm{Adj}(A)B]+\mathcal{O}(B^2),
\end{equation}
with $\mathrm{Adj}(A)$ denoting adjugate matrix of $A$, we get

\begin{equation}
    C_{ij}^{xx}= (-1)^{j-i+1}F_{ij}+\mathcal{O}(F_{kl}^2)=\pm \frac{1}{2}\sqrt{|C_{ij}^{zz}|}+\mathcal{O}\left(|C^{zz}_{kl}|\right).
\end{equation}
In the presence of interactions at moderate value of disorder, corrections in above formula become significant, and relation~\eqref{eq:xi_x_vs_xi_z} is not satisfied.

\section{Comparison of spectral measures}

\begin{figure}[t]
    \centering
    \includegraphics[width=.75\linewidth]{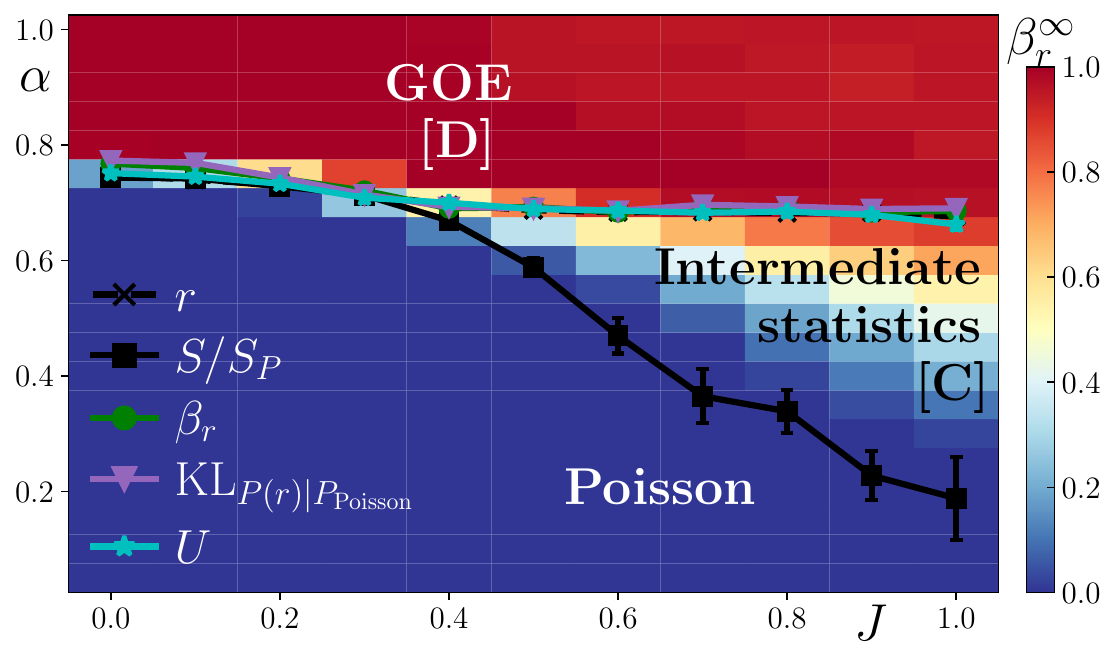}
    \caption{Comparison of phase transitions for eigenstate-based entanglement entropy measure $S/S_P$ and eigenvalue-based measures: mean gap ratio $r$, the coefficient $\beta_r$ in \eqref{eq:pr}, Kullback–Leibler divergence in \eqref{eq:kl} and Binder cumulant $U$ in \eqref{eq:binder}. Color map corresponds to extracted value of $\beta_r^\infty$ in thermodynamic limit, for given $\alpha$ and $J$.}
    \label{fig:spectral_measure_comparison}
\end{figure}

To complement the analysis of spectral properties of the model, we compare average spectral gap ratio $\langle r\rangle$ with other energy-based measures. We consider Dyson index $\beta_r\in[0,1]$, which we extract from the fit of the distribution $P(r)$ to Wigner-Dyson statistics

\begin{equation}
    P(r) \propto \frac{(r + r^2)^{\beta_r}}{(1 + r + r^2)^{1 + 3\beta_r/2}},
    \label{eq:pr}
\end{equation}
where $\beta_r=0$ corresponds to Poisson statistics, while the GOE has $\beta_r=1$. Dyson index $\beta_r$ contains information about the whole distribution, beyond just the average $\langle r \rangle$. We also compute the Kullback–Leibler divergence to the Poisson distribution

\begin{equation}
    \mathrm{KL}_{P(r)|P_{\rm Poisson}}=\int_0^1  P(r)\ln\frac{P(r)}{P_{\rm Poisson}(r)}\mathrm{d}r,
    \label{eq:kl}
\end{equation}
with the limiting values $\mathrm{KL}_{P_{\rm Poisson}|P_{\rm Poisson}}=0$ and $\mathrm{KL}_{P_{\rm GOE}|P_{\rm Poisson}}=1/3+\ln(\sqrt{3}/2)\approx0.18949$. Expanding $\ln \frac{P(r)}{P_{\rm Poisson}(r)}$ in $r$, it is clear that $\mathrm{KL}_{P(r)|P_{\rm Poisson}}$ contains information on all moments of $P(r)$. Finally, we consider a fourth order cumulant of $P(r)$ called the Binder cumulant $U$

\begin{equation}
    U=1-\frac{\langle r^4\rangle}{3\langle r^2\rangle^2},
    \label{eq:binder}
\end{equation}
which for the Poisson ensemble is $U=\frac{(\ln2-2/3)^2}{(\ln2-3/4)^2}\approx0.21695$, whereas GOE has $U=1-\frac{938+135\ln(2\sqrt{3}-3)-478\sqrt{3}}{6[2+27\ln(2\sqrt{3}-3)+10\sqrt{3}]^2}\approx 0.45655$.

For all of those spectral measures, we repeat the same procedure as we did for the gap ratio $r$, extracting the crossings for $J\leq0.2$ or thresholds close to GOE limit for $J\geq 0.3$, with $\beta_{r,\rm thresh}=0.16$, $\mathrm{KL}_{\rm thresh}=0.025$, $U_{\rm thresh}=0.03$. As a result of this procedure, we get a prediction for a critical point of the model in the thermodynamic limit $\alpha_\infty^{\rm c}$ for each of those measures, see Fig.~\ref{fig:spectral_measure_comparison}. All the measures considered agree well within the error bars, confirming the robustness of the predicted transition between phases~[C] and~[D]. Moreover, extrapolations suggest that phase~[C] takes intermediate values of $\beta_r^\infty$, while the value of the Poisson ensemble is reached only in the region~[B] and~[A], which corresponds to a region of the phase diagram below the solid black $S/S_P$ line in Fig.~\ref{fig:spectral_measure_comparison}. This behavior is identical to that of the gap ratio $r$ shown in Fig.~\ref{fig:phase_diagram}(a) of the main text, and it is also shared by the remaining spectral measures, namely the Kullback–Leibler divergence and the Binder cumulant $U$.

\section{Scaling of entanglement entropy}

In order to demonstrate that region~[B] of the phase diagram has intermediate scaling of the entanglement entropy, between area-law and volume-law, we perform a fit $S\propto L^b$ for fixed $\alpha,J$. To account for finite-size effects, we perform this fit limiting the system sizes to a window $[L-2,L+2]$, obtaining different $b$ for each choice of the window center $L$.

\begin{figure}[ht]
    \centering
    \begin{subfigure}[b]{0.48\textwidth}
        \centering
        \includegraphics[width=\linewidth]{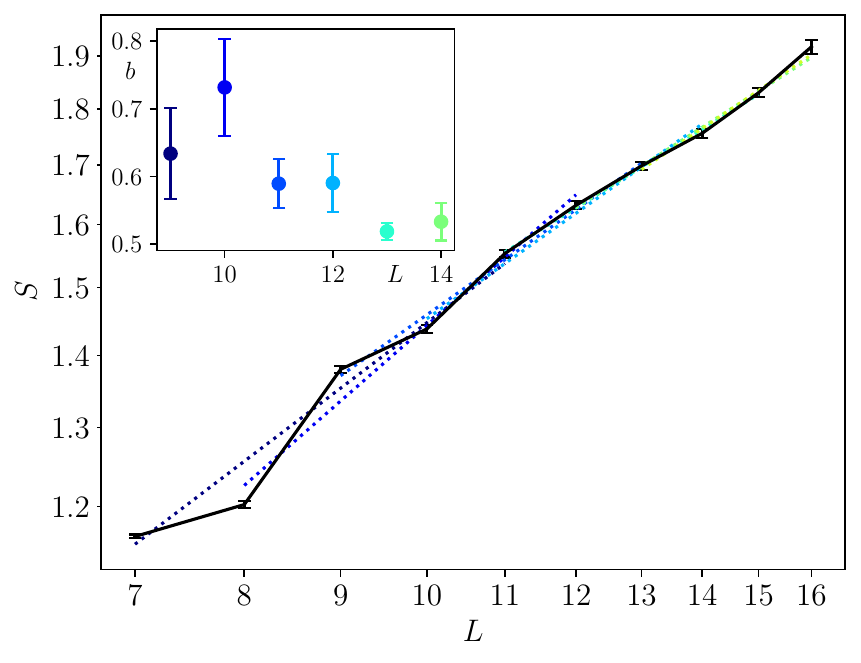}
        \caption{$J=0.6,\alpha=0.3$, phase [B].}
    \end{subfigure}
    \hfill
    \begin{subfigure}[b]{0.48\textwidth}
        \centering
        \includegraphics[width=\linewidth]{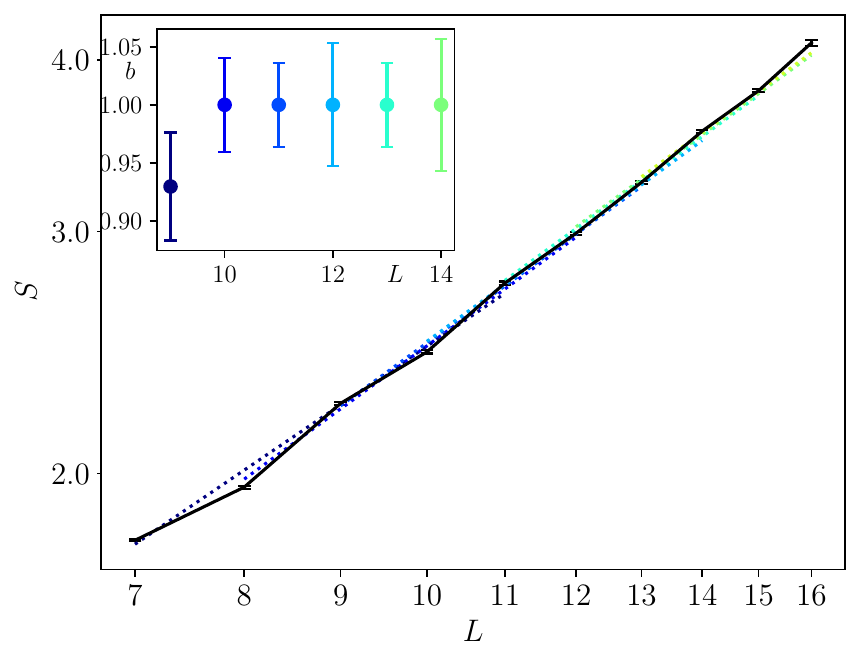}
        \caption{$J=0.6,\alpha=0.6$, phase [C]}
    \end{subfigure}

    \caption{Dependence of averaged entanglement entropy on system size $L$ on log-log scale. Fits $S=aL^b$ to system sizes $[L-2,L+2]$ are shown with dotted lines. Obtained values of $b$ are shown in the inset, with colors corresponding to fits.}
    \label{fig:S_vs_L}
\end{figure}

In Fig.~\ref{fig:S_vs_L}, we present the $S(L)$ dependence for two representative points from regions~[B] and~[C], together with the extracted values of $b$ in insets. In regime~[B] robust power-law dependence of $S(L)$ is observed, with $b\in(0,1)$, whereas in~[C], which is characterized by intermediate statistics, the value $b=1$ is quickly reached as the window center $L$ is increased.

\begin{figure}[t]
    \centering
    \begin{subfigure}[b]{0.48\textwidth}
        \centering
        \includegraphics[width=\linewidth]{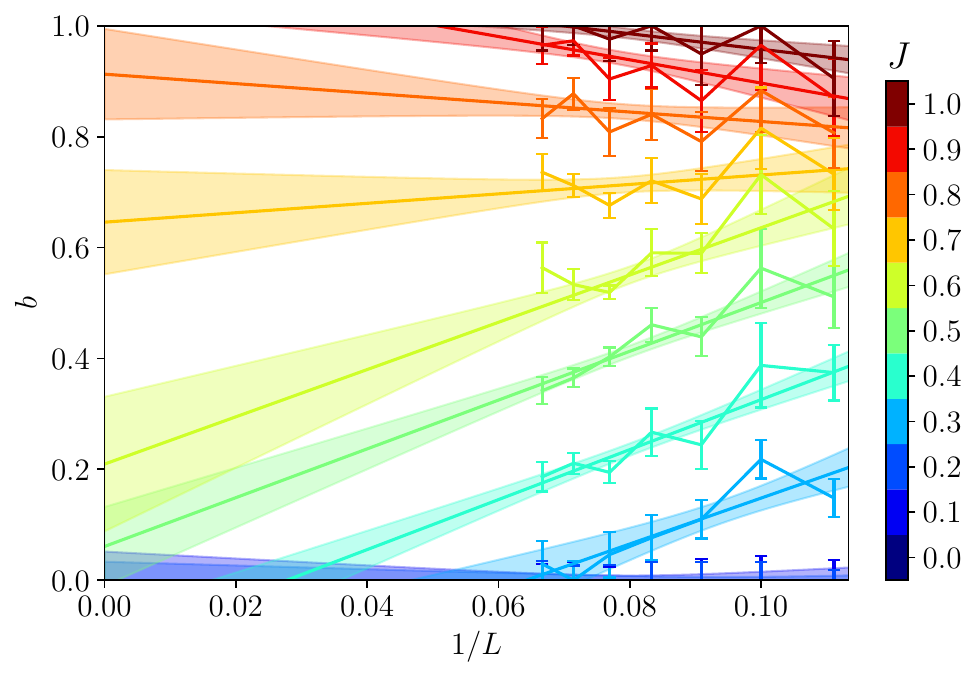}
        \caption{$\alpha=0.3$.}
    \end{subfigure}
    \hfill
    \begin{subfigure}[b]{0.48\textwidth}
        \centering
        \includegraphics[width=\linewidth]{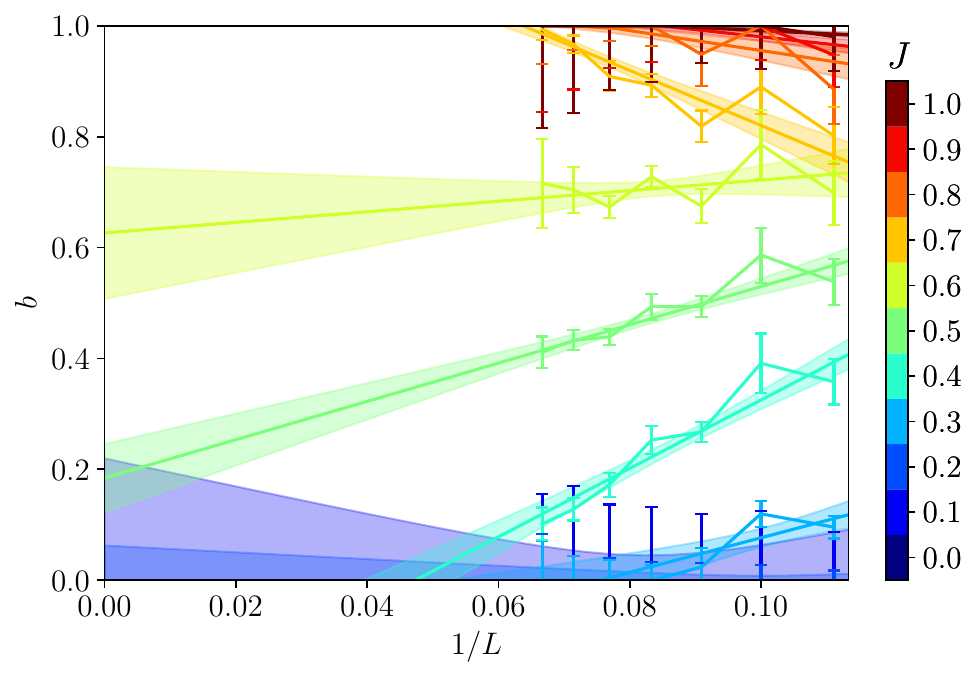}
        \caption{$\alpha=0.4$}
    \end{subfigure}

    \caption{Finite-size scaling of power $b$ in entanglement entropy dependence $S\propto L^b$. For given $L$, parameter $b$ is extracted from the fit to system sizes $[L-2,L+2]$.}
    \label{fig:b_vs_L}
\end{figure}

We extract the power in the thermodynamic limit $b^\infty$, using the fact that $b(L)$ is well described by the fit $b(L)=b^\infty+b_1/L$. The value $b^{\infty}$ extracted this way is used as a colormap in Fig.~\ref{fig:phase_diagram}(b) of the main text. In Fig.~\ref{fig:b_vs_L} we present those extrapolations for $\alpha=0.3$ and $\alpha=0.4$ and all studied values of $J=0,\,0.1,\,\ldots,\,1$. This choice of the horizontal cut on the phase diagram goes through regimes~[A],~[B] and~[C]. Those regions can be distinguished from each other with the extracted value of $b^{\infty}$.

\end{document}

%% file: arxiv.bbl
%